\documentstyle[aps,twocolumn]{revtex}
\begin{document}

\draft

\title{Trapped ions in the strong excitation regime:
ion interferometry and non--classical states}

\author{J. F. Poyatos, J. I. Cirac}
\address{Departamento de Fisica Aplicada, Universidad de
Castilla--La Mancha, 13071 Ciudad Real, Spain}

\author{R. Blatt}
\address{Institut f\"ur Experimental Physik, Universit\"at
G\"ottingen, 37073 G\"ottingen, Germany}

\author{P. Zoller}
\address{Institut f{\"u}r Theoretische Physik,
Universit{\"a}t Innsbruck, 6020
Innsbruck, Austria}

\date{\today}

\maketitle

\begin{abstract}
The interaction of a trapped ion with a laser beam in the
strong excitation regime is analyzed. In this regime, a variety
of non--classical states of motion can be prepared either by
using laser pulses of well defined area, or by an adiabatic
passage scheme based on the variation of the laser frequency.
We show how these states can be used to investigate fundamental
properties of quantum mechanics. We also study possible
applications of this system to build an ion interferometer.
\end{abstract}


\narrowtext


\section{Introduction}

In recent years, there has been a growing interest in the
preparation of non--classical states of quantum systems in
order to study fundamental properties of quantum mechanics. In
the cavity QED context \cite{cQEDreview} for example, there are
several proposals to prepare Fock states of the radiation field
\cite{Fi86}, or general superpositions of these states
\cite{Pa93}. For single trapped ions, it has been shown how to
prepare both Fock and squeezed states of the motion of an ion
using a laser beam to excite an internal transition
\cite{Ci95a}.

{}From the experimental point of view, the main difficulty to
prepare non--classical states in a given system is the presence
of decoherence due to its coupling to external environments.
In cavity QED this may be overcome by using high--Q cavities,
where the coupling strength between the atoms and the cavity
mode is of the order of or larger than both, the cavity loss
rate and the spontaneous emission rate. With trapped ions, one
can use an electric dipole forbidden transition in order to
avoid dissipation. In this last case, the quantum jump
technique \cite{Na86} allows one to peform measurements on the
internal atomic state with a very high efficiency.

So far, all the proposals regarding the preparation of
non--classical states of motion of a single ion operate in the
{\it low excitation regime}, whereby the Rabi frequency
corresponding to the interaction of the ion with the laser
$\Omega$ is much smaller than the trap frequency $\nu$ ($\Omega \ll
\nu$). In this case, the Hamiltonian describing this
interaction is very similar to the one that describes cavity
QED, namely the Jaynes--Cummings Hamiltonian. Thus, some of the
quantum phenomena predicted in cavity QED, such as collapse and
revival or vacuum Rabi splitting, have been predicted to exist
for trapped ions under this regime \cite{Bl92,Ci93a}.
Furthermore, the unique features of the ion--laser
interaction has led to various proposals on how to prepare
non--classical states of motion, which have no analog in cavity
QED. For example, Fock states can be prepared by observing
quantum jumps in the internal state of the ion \cite{Ci93a}, or
by sending short laser pulses with well defined area to the ion
\cite{Ci95b}. Coherent and squeezed states of motion can be
prepared by bichromatic excitation of the ion \cite{Ci93b}.

In this paper we analyze the possibility of preparing
non--classical states of motion of a single ion in a completely
different regime. We concentrate on the {\it strong excitation
regime}, whereby the Rabi frequency describing the interaction
of the ion with the laser is much larger than the trap
frequency ($\Omega \gg \nu$). In this regime, the Hamiltonian
describing such interaction is completely different from the
one used in cavity QED, and therefore other novel phenomena can
be investigated. In particular, we will show here how one can
prepare quantum superpositions of two ``macroscopically'' distinct
quantum states, such as
\begin{equation}
\label{Schcatint}
|\Psi\rangle = K (|\alpha\rangle_c + |-\alpha\rangle_c),
\end{equation}
where $K$ is a normalization constant, $|\alpha\rangle_c$
is a coherent state of the motion
\begin{equation}
\label{cohstate}
|\alpha\rangle_c = e^{-|\alpha|^2/2} \sum_{n=0}^\infty
\frac{\alpha^n}{\sqrt{n!}} |n\rangle,
\end{equation}
and $|n\rangle$ denotes a Fock state with $n$ phonons
\cite{note1}. States (\ref{Schcatint}) are usually called
Schr\"odinger cat states \cite{Sc35,Yu86}. The fundamental
properties of quantum mechanical superpositions of states such
as (\ref{Schcatint}) have attracted due attention, and many
schemes for their production have been proposed. In the cavity
QED context, it has been shown \cite{Ge90} that during the
collapse time, the state of the field is of the form
(\ref{Schcatint}). On the other hand, with a micromaser
interaction one can also produce these states by sending atoms
through a cavity \cite{Br92}. The method we propose consists of
two parts: first, using a laser one splits into two parts the
ion wavefunction; then, one makes a measurement of the {\it
internal} state of the ion in order to project its {\it
external} state into (\ref{Schcatint}). Thus, the projection
postulate of quantum mechanics is one of the ingredients of our
method. We will show two different ways to prepare the state
(\ref{Schcatint}), one based on laser pulses and the other on
the adiabatic change of the laser frequency during the
interaction \cite{Ci94}. We will also show how to prepare
linear superpositions of coherent states in more than one
dimension, and how to distinguish between that state and an
incoherent superposition of two coherent states. Finally, we
will study how a Ramsey interferometer \cite{Ramseyint} can be
built using these ideas.

This paper is organized as follows: first, in Section II we
give a description of the strong excitation regime, and find
analytical approximations for the evolution of a single ion
under this regime. In Section III we show how to prepare states
of the form (\ref{Schcatint}) in one and two dimensions, and in
Section III we analyze how these states can be distinguished from
statistical mixtures. The possibility of building an
interferometer with trapped ions is analyzed in Section IV. A
summary of the results of the paper is given in Section V.

\section{Strong excitation regime}

In this Section we analyze the interaction of a trapped ion
with a laser beam in the strong excitation regime. In this
regime, the Rabi frequency for the laser--ion interaction is
much larger than the trap frequency ($\Omega \gg \nu$), i.e. one
can neglect the ion motion during the interaction. We will also
assume that the laser is tuned to an electric dipole forbidden
transition $|g\rangle \leftrightarrow |e\rangle$ (see Fig.~1),
and therefore we will neglect spontaneous emission. Under
these conditions the problem becomes exactly solvable. Here we
will study how the motion is modified after this interaction in
two different situations: (i) the ion interacts with a laser
pulse of well defined area; (ii) the frequency of the laser is
varied adiabatically.

\subsection{Excitation with laser pulses}

Let us consider a single ion trapped in a three dimensional
harmonic potential. The ion interacts with a laser plane wave
propagating along the $x$--axis. In this configuration, only
the motion along the $x$ axis of the ion will be modified, so
that we can treat this problem in one--dimension. The
Hamiltonian describing this situation is, in a rotating frame
at the laser frequency $\omega_L$ ($\hbar=1$),
\begin{equation}
\label{H_-}
H_\pm = \frac{\hat p_x^2}{2m} + \frac{1}{2} m \nu_x^2 \hat x^2
- \frac{\delta}{2} \sigma_z
+ \frac{\Omega}{2} \left(
  \sigma_+ e^{\pm i k_L \hat x} + \sigma_- e^{\mp i k_L \hat x}
  \right).
\end{equation}
Here, $\hat x$ and $\hat p_x$ are the position and momentum
operators for the $x$ coordinate of the ion, $\nu_x$ is the
trap frequency along this direction, and $m$ is the ion mass.
The sigmas are usual spin $\frac{1}{2}$ operators describing
the internal two--level transition $|g\rangle \leftrightarrow
|e\rangle$, $\delta=\omega_L-\omega_0$ the laser detuning,
$k_L=\omega_L/c$ the laser wavevector, and $\Omega$ the Rabi
frequency. The subscript ``$+$'' (``$-$'') indicate that the
laser plane wave propagates towards the positive
(negative) values of $x$.

The evolution given by Hamiltonian (\ref{H_-}) is easily
described if we perform a unitary operation defined by
\begin{equation}
\label{U_-}
U_\pm = e^{\mp i k_L \hat x |e\rangle \langle e|}.
\end{equation}
Using this operator, the states are defined as
\begin{equation}
|\tilde \Psi_\pm \rangle = U_\pm |\Psi \rangle,
\end{equation}
whereas the Hamiltonian becomes
\begin{eqnarray}
\label{tildeH_-}
\tilde H_\pm \equiv U_\pm H_\pm U_\pm^\dagger
& &= \nu_x a^\dagger a
\pm i \nu_x \eta_x (a-a^\dagger) |e\rangle \langle e| \nonumber\\
& &+ \nu_x \eta_x^2 |e\rangle \langle e|
- \frac{\delta}{2} \sigma_z + \frac{\Omega}{2} \sigma_x,
\end{eqnarray}
where $\sigma_x=\sigma_++\sigma_-$, and we have expressed the position and
momentum operators in terms of the annihilation and creation
operators of the harmonic oscillator, $a$ and $a^\dagger$,
respectively, and in terms of the Lamb--Dicke parameter
$\eta_x= k_L /(2m\nu_x)^{1/2}$. The new terms appearing in
Hamiltonian (\ref{tildeH_-}) correspond to the Doppler recoil
energy of the excited internal state (i.e., when one photon is
absorbed).

Hamiltonian (\ref{tildeH_-}) can be simplified in the
strong excitation regime $\Omega \gg \nu_x$. Let us consider a
laser pulse whose duration $\tau$ fulfills the following
inequality
\begin{equation}
\label{condnut}
\nu_x \tau \; {\rm max} ({\overline n},\eta_x^2)
\ll 1,
\end{equation}
where ${\overline n} = \langle a^\dagger a \rangle$. Note that in the
strong excitation limit, this interaction time can correspond
to pulses with an area $\Omega \tau \sim \pi$. Under
condition (\ref{condnut}), the first terms in Hamiltonian
(\ref{tildeH_-}) will practically not affect the evolution. In
this case, and assuming for simplicity that the detuning is
zero ($\delta=0$), the Hamiltonian reduces to $\tilde H =
\frac{1}{2} \Omega \sigma_x$ (for both cases, $\tilde H_\pm$)
\cite{note2}. The evolution of any initial state can be easily
derived, obtaining
\begin{equation}
\label{evolgen}
|\Psi_\pm(\tau)\rangle =
U_\pm^\dagger e^{-i \frac{1}{2} \Omega \sigma_x \tau} U_\pm
|\Psi(0)\rangle.
\end{equation}
Note that the evolution given by $H_-$ can be obtained from
that given by $H_+$ by simply exchanging $\eta_x
\rightarrow -\eta_x$.

In the following we will need to know the evolution for the
case where the initial state of motion along the $x$ direction
is a coherent state $|\alpha\rangle_c$. In this situation,
Eq.~(\ref{evolgen}) gives
\begin{mathletters}
\label{evolpuls}
\begin{eqnarray}
|g\rangle |\alpha\rangle_c &\rightarrow&
A |g\rangle |\alpha\rangle_c
+ B |e\rangle |\alpha \pm i\eta_x \rangle_c, \\
|e\rangle |\alpha\rangle_c &\rightarrow&
A^\ast |e\rangle |\alpha\rangle_c
-B^\ast |g\rangle |\alpha \mp i\eta_x \rangle_c,
\end{eqnarray}
\end{mathletters}
where the upper (lower) sign corresponds to a laser pulse
propagating towards the positive (negative) values of the $x$
axis. For reasons that will become clear in the next
subsection, we have defined in (\ref{evolpuls})
\begin{mathletters}
\begin{eqnarray}
A &=& \cos(\Omega \tau/2),\\
B &=& -i \sin(\Omega \tau/2).
\end{eqnarray}
\end{mathletters}

Finally, in this Section we have assumed square laser pulses.
For any other kind of pulse, formulas (\ref{evolpuls}) are
still valid, but now one has to replace $\Omega \tau$ by
$\int_0^\tau \Omega(t) dt$.

\subsection{Excitation by adiabatic passage}

In this subsection we will analyze a different method to obtain
results that are similar to those of the previous subsection.
Here, instead of using a laser pulse to modify the internal and
external state of the ion, we will consider a process in which
the laser frequency is changed adiabatically.

Let us consider then the same situation as before. A trapped
ion interacts with a laser plane wave propagating along the $x$
direction. The Hamiltonian describing this situation, after
applying the unitary operation (\ref{U_-}) is given by
Eq.~(\ref{tildeH_-}), although now it is time dependent
(through the time dependence of the detuning). As before, for
short interaction times $\tau$ fulfilling Eq.~(\ref{condnut}),
the Hamiltonian can be simplified to
\begin{equation}
\tilde H(t) =
- \frac{\delta(t)}{2} \sigma_z
+ \frac{\Omega}{2} \sigma_x.
\end{equation}
At a given time $t$, the instantaneous eigenstates of this
Hamiltonian are the well--known dressed states
\begin{mathletters}
\begin{eqnarray}
|+(t)\rangle &=& \cos[\theta(t)] |e\rangle
+ \sin[\theta(t)] |g\rangle, \\
|-(t)\rangle &=& -\sin[\theta(t)] |e\rangle
+ \cos[\theta(t)] |g\rangle,
\end{eqnarray}
\end{mathletters}
with corresponding eigenvalues
\begin{equation}
E_{\pm}(t) = \pm \frac{1}{2} \sqrt{\delta(t)^2 + \Omega^2},
\end{equation}
and where cot$[2\theta(t)]=-\delta(t)/\Omega$ ($0 \le 2\theta <
\pi$).

By changing the detuning adiabatically from
$\delta_0=\delta(0)$ to $\delta_\tau=\delta(\tau)$ the internal
state of the ion will follow (approximately) the evolution of
these dressed states, i.e.,
\begin{equation}
\label{dressevol}
|\pm(0)\rangle \rightarrow
e^{\mp i \epsilon} |\pm(\tau)\rangle,
\end{equation}
where
\begin{equation}
\epsilon=\int_0^\tau E_+ (t) dt.
\end{equation}

The condition for the adiabatic passage to be valid
(\ref{dressevol}) can be easily estimated \cite{Me65}. For
example, assuming that the rate of change of $\delta$ is
constant, one finds
\begin{equation}
\label{condadpas}
|\delta_0-\delta_\tau | \ll \Omega^2 \tau.
\end{equation}

Using Eqs.~(\ref{U_-}) and (\ref{dressevol}), we find
(\ref{evolpuls}) for the evolution of an ion initially in a coherent
state of motion, where now
\begin{mathletters}
\label{ABadpas}
\begin{eqnarray}
A &=&
\cos(\epsilon)\cos(\theta_0-\theta_\tau)
+i \sin(\epsilon)\cos(\theta_0+\theta_\tau),\\
B &=&
\cos(\epsilon)\sin(\theta_0-\theta_\tau)
-i \sin(\epsilon)\sin(\theta_0+\theta_\tau),
\end{eqnarray}
\end{mathletters}
and $\theta_0=\theta(0)$ [$\theta_\tau=\theta(\tau)$]. Thus, by
choosing appropriately the initial and final detunings and
provided the adiabatic condition (\ref{condadpas}) is fulfilled
one can achieve by adiabatic passage the same effect as using
laser pulses.

\section{Superpositions of coherent states}

In this Section we show how superpositions of coherent states
(Schr\"odinger cats) of the ion motion can be prepared using a
laser beam in the strong excitation regime. We will analyze
both the one--dimensional and the two--dimensional cases.

\subsection{1--Dimension}

Let us assume that after sideband cooling \cite{Di89}, we have
the ion in the ground state of both, the internal and external
degrees of freedom ($|g\rangle |0\rangle$). In this subsection
we discuss two ways of preparing states of the form
(\ref{Schcatint}) in one--dimension ($x$), one based on laser
pulses and the other on adiabatic passage by varying the laser
frequency. For the sake of a simple notation, we will drop the
subscript $x$ in all the states and formulas when we deal with
one--dimensional problems.

\subsubsection{Pulses}

A simple way to prepare the state (\ref{Schcatint}) using
laser pulses consists of the following three steps:

\begin{description}

\item (i) Excite the ion with a $\pi/2$ pulse ($\Omega \tau= \pi/2$)
using a laser propagating towards the negative values of the
$x$ axis. According to (\ref{evolpuls}), this pulse will perform
the following transformation
\begin{equation}
|g\rangle |0\rangle_c \rightarrow
\frac{1}{\sqrt{2}}
(|g\rangle |0\rangle_c - i |e\rangle |-i\eta\rangle_c),
\end{equation}
where $|\alpha\rangle_c$ denotes the coherent state
(\ref{cohstate}).

\item (ii) Excite the ion with another $\pi/2$ pulse
($\Omega \tau= \pi/2$), but now using a laser beam propagating in
the opposite direction. The state of the ion will become
\begin{equation}
\label{step2}
\frac{1}{2}
[ (|0\rangle_c - |-2i\eta\rangle_c) |g\rangle
- i (|i\eta\rangle_c + |-i\eta\rangle_c) |e\rangle ].
\end{equation}

\item (iii) Measure the internal state of the ion using the
quantum jump technique \cite{Na86}. To do this one can drive
the ion with a different laser beam on resonance with an
electric dipole allowed transition $|g\rangle \leftrightarrow
|r\rangle$ (see Fig.~1). In case fluorescence is observed, the
state of the ion will be projected onto the part of the
wavefunction (\ref{step2}) that contains the state $|g\rangle$.
On the contrary, if no fluorescence is observed, it will be
projected onto the state
\begin{equation}
\label{Schcat}
|\Psi_{sc}\rangle = K (|i\eta\rangle_c +
|-i\eta\rangle_c)|e\rangle,
\end{equation}
where $K$ is a normalization constant.

\end{description}

State (\ref{Schcat}) is already of the form (\ref{Schcatint}),
i.e. similar to that studied in cavity QED
\cite{Ge90,Br92}(whereby the motion of the ion corresponds to
the cavity mode). Note that if fluorescence is observed, the
external state of the motion will be completely modified due to
the photon recoil acquired by the ion in each absorption
spontaneous emission cycle, and therefore no such a
superposition of coherent states will be produced. Therefore,
an experiment based on these steps will be successful half of
the times it is carried out (those in which no fluorescence is
observed).

On the other hand, in order to consider state (\ref{Schcat}) as
a Schr\"odinger cat it should be macroscopic (in the sense that
one can observe both states $|\alpha\rangle$ and $|-\alpha
\rangle$ individually \cite{note1}). However, there are two
reasons why the state (\ref{Schcat}) cannot be directly
observable as it stands: (a) its probability distribution in
position representation has only one peak centered at $\langle
\hat x \rangle =0$, and therefore the two parts of such state
($|i\eta\rangle$ and $|-i\eta\rangle$) cannot be distinguished
by direct observation; (b) Since the ion is in its internal
excited state, it can not be observed by shining light on the
$|g\rangle \leftrightarrow |r\rangle$ transition. This second
problem (b) can be easily solved if, just after the state
(\ref{Schcatint})  is produced, one applies a laser $\pi$ pulse
along the $z$ direction, that transforms the excited state into
the ground state. Note in order not to modify the state of the
ion with this last pulse it is necessary for the ion to be
confined in the Lamb--Dicke limit in the $z$ direction, i.e.
$\eta_z=k_L /(2m\nu_z)^{1/2} \ll 1$. This may be the case, for
example, in a linear ion trap,
whereby the transverse directions have trap frequencies much
larger than along the axial direction \cite{Ra92}". The first problem
(a) can be solved by noting that the free evolution of a
coherent state is given by $|\alpha(t)\rangle = |e^{-i\nu t}
\alpha(0)\rangle$ and, therefore, if after producing the state
(\ref{Schcatint}), one waits for a time $t=\pi/(2\nu)$, the
state will become
\begin{equation}
\label{Schcatp}
|\Psi_{sc}\rangle = K (|\eta\rangle_c +
|-\eta\rangle_c)|e\rangle.
\end{equation}
This state has two maxima (in position representation) centered
at $\langle \hat x \rangle=\pm 2 \eta /\sqrt{2m\nu}$,
respectively (each of these maxima correspond to the states
$|\pm \eta\rangle$, respectively). Furthermore, in order to be
able to observe these two states the corresponding peaks must
be spatially separated by more than a wavelength, which
requires $\eta^2 > \pi/2$. For tight traps, that is, for values
of $\eta$ not fulfilling this condition, one can proceed as
follows. After the step (i), one applies a sequence of $\pi$
pulses from the left and from the right ($n$ pulses in each
direction), in an alternating way \cite{We94}. It is easy to
check that the state after step (iii) and free evolution
will be
\begin{equation}
\label{Schcat2}
|\Psi_{sc}\rangle = K (|(2n+1)\eta\rangle_c +
|-(2n+1)\eta\rangle_c)|e\rangle.
\end{equation}
For these states, the number of pulses required to observe
distinguishible (macroscopic) states is $(2n+1)^2  >
\pi/(2\eta^2)$. Note also that now in condition
(\ref{condnut}) it is the total time corresponding to all the
pulses the one that enters.

In order to illustrate the effectiveness of the method
presented here, we have plotted in Fig.~2 the state after step
(iii) for several values of $\Omega/\nu$ and $\eta$. To produce
these plots, we have solved numerically the evolution equations
of the ion using the exact Hamiltonian (\ref{H_-}). The figures
display the real part of the density operator in momentum
representation $\langle p|\rho|p'\rangle$ (note that the axes
are rescaled in terms of $p_0=\sqrt{m\nu/2}$). Figures 2(a,b,c)
correspond to $\eta=0.5$ and $n=2$ (i.e. two intermediate $\pi$
pulses in each direction), whereas Figs.~2(d,e,f) correspond to
$n=0$ with $\eta=2.5$. For $\Omega/\nu=100$ [Figs.~2(a,d)] the
method works almost ideally, since condition (\ref{condnut}) is
satisfied. There are four peaks in the plots, two of them
corresponding to the diagonal parts of the density operator
($|i\eta\rangle \langle i\eta|$ and $|-i\eta\rangle \langle
-i\eta|$), and the other corresponding to the coherences
($|-i\eta\rangle \langle i\eta|$ and $|i\eta\rangle \langle
-i\eta|$). These last two peaks are the ones ensuring that the
state is a truly pure state, as opposed to a statistical mixture
in which these two peaks will not show up. As soon as the ratio
$\Omega/\nu$ is decreased, so that condition (\ref{condnut}) is
not satisfied, this four peak structure disappears. Note that the
plots corresponding to $n=2$ are much more sensitive to this
condition than those with $n=0$. On the other hand, we have
checked that they are more robust with respect to mismatches
in the pulse areas as well.

\subsubsection{Adiabatic passage}

Instead of using controlled laser pulses, one can use the
adiabatic passage technique to produce the superposition of
coherent states. This technique has the advantage that it is
not sensitive to the specific values of the parameters
characterizing the laser--ion interaction (such as interaction
time, Rabi frequencies, etc). The process is very similar to
the one explained above for pulses. It consists of three steps:

\begin{description}

\item (i) A laser beam propagating towards the negative values
of the $x$--axis is directed to the ion. The detuning is
switched adiabatically from $\delta_0= - \Delta$ to
$\delta_\tau = 0$, with $\Delta \gg \Omega$. In this case,
according to (\ref{ABadpas}) the state of the ion will become
($\theta_0=0,\theta_\tau=\pi/4$)
\begin{equation}
|g\rangle |0\rangle_c \rightarrow
\frac{1}{\sqrt{2}}
(|g\rangle |0\rangle_c - |e\rangle |-i\eta\rangle_c).
\end{equation}

\item (ii)  A laser beam propagating towards the positive
values of $x$ is directed to the ion. The detuning is switched
adiabatically from $\delta_0= 0$ to $\delta_\tau = \Delta$,
again with $\Delta \gg \Omega$. The state of the ion after this
step will be ($\theta_0=\pi/4,\theta_1=\pi/2$)
\begin{equation}
\label{step2p}
\frac{1}{2}
[e^{-i\epsilon} (|0\rangle_c - |-2i\eta\rangle_c) |g\rangle
- e^{i\epsilon} (|i\eta\rangle_c + |-i\eta\rangle_c) |e\rangle ].
\end{equation}

\item (iii) Measure the internal state of the ion using the
quantum jump technique \cite{Na86}. If no fluorescence is
observed, the external state of the ion will be projected onto
\begin{equation}
\label{Schcatpp}
|\Psi_{sc}\rangle = K (|i\eta\rangle_c +
|-i\eta\rangle_c)|e\rangle.
\end{equation}

\end{description}

As before, one can include intermediate steps between (i) and
(ii) to achieve a larger extension between the peaks of the
probability distribution. To this aim, one can perform sequences of
excitations, using counterpropagating lasers: first, changing
the detuning from $\delta_0=0$ to $\delta_1=\Delta$, and then
from $\delta_0=-\Delta$ to $\delta_1=0$. By performing this
operation $n$ times (in each direction), the final state of the
ion (after measurement) will be
\begin{equation}
\label{Schcatppp}
|\Psi_{sc}\rangle = K (|(2n+1)i\eta\rangle_c +
|-(2n+1)i\eta\rangle_c)|e\rangle.
\end{equation}

In Fig.~3 we have plotted the real part of $\langle p'|\rho| p
\rangle$ for the state arising after step (iii), by solving
numerically the evolution equations using the full Hamiltonian
(\ref{H_-}). Here, $\eta=0.5$, and the number of intermediate
steps is $n=2$. We have taken $\Delta=10\Omega$ and
$\Omega/\nu=100$. The times for each adiabatic passage step are
$\tau = 40,50,60$ and $70\Omega^{-1}$ [Figs.~3(a,b,c,d),
respectively]. Note that with these parameters the conditions
of no motion during the interaction (\ref{condnut}) and of
adiabatic passage (\ref{condadpas}) are nearly fulfilled. By
comparing these four plots, one sees that in all of them four
peaks show up, corresponding to the diagonal part and the
coherences. However, in some of the plots there is an extra
wide peak in the center, which does not disappear even if the
interaction time is taken to be longer (in order to improve the
adiabatic passage requirements). In fact, we have checked
numerically that the peak appears nearly periodically as the
time $\tau$ is increased. The reason for this peak is related
to the fact that $\Delta$ is finite in the plots. Using
relations (\ref{ABadpas}) it can be easily shown that the four
peaks (in $\langle p'|\rho| p \rangle$)  oscillate due to the
dynamical phase $\epsilon$, and the amplitude of the
oscillations is of the order of $n \Omega/\Delta$, where $n$ is
the number of itermediate steps. The best results are obtained
for $n=0$, where there is no such oscillation; but this
requires a large value for $\eta$ in order for the two
locations of the ion to be observable. This may be realized
with a linear ion trap, for which the motion along the axial
direction can have $\eta >1$. Summarizing, in order to
prepare a state of the form (\ref{Schcatint}) using adiabatic
passage with intermediate laser pulses, apart from conditions
(\ref{condnut}) and (\ref{condadpas}), it is required that
$\Delta\gg\Omega$. Note that this last condition is somehow
difficult to achieve experimentally, since as one increases
$\Delta$, a longer time is required to fulfill the adiabatic
condition, which may require a very small trap
frequency in order not to violate condition (\ref{condnut}).

\subsection{2--Dimensions}

In this subsection we generalize the above methods to show how
superpositions of coherent states in 2 or more dimensions can
be generated. The basic idea is to prepare first a
superposition state along a given direction following the steps
presented in the previous subsections, and afterwards to send
pulses along a perpendicular direction in order to push the
wavepackets and obtain a circular motion. The steps required to
implement a superposition of coherent states in the $x-y$
dimensions are the following:

\begin{description}

\item (i) After sideband cooling \cite{Di89}, prepare a the
state (\ref{Schcatint}) in the $x$ direction as indicated in
the previous subsection. Then wait for a time
$\tau=\pi/(2\nu)$. The state of the ion will be
\begin{equation}
|\Psi_{sc}\rangle = K (|\eta\rangle_x +
|-\eta\rangle_x) |0\rangle_y |e\rangle.
\end{equation}

\item (ii) Drive the ion with a $\pi$ pulse using a laser
propagating along the $y$ direction. The state will become
\begin{equation}
\label{Schcat2d}
|\Psi_{sc}\rangle = K (|\eta\rangle_x +
|-\eta\rangle_x) |-i\eta\rangle_y |g\rangle.
\end{equation}

\end{description}

State (\ref{Schcat2d}) is composed of two coherent wavepackets
each of them propagating along a circle in the $x-y$ plane
(provided the trap frequencies $\nu_x=\nu_y$), but in
opposite directions. Note that the internal state of the ion
after the pulse sequence is $|g\rangle$, and therefore it is
ready to be detected by quantum jumps. Obviously, for a trap
with a small value of $\eta$ one can add intermediate $\pi$ pulses in
both directions $x$ and $y$ in order to make a larger circle
radius. On the other hand, one can use the adiabatic passage
technique instead of using pulses, utilizing the same procedure.

In Fig.~4 we have plotted snap--shots of the probability
distribution for the ion in position representation
$P(x,y)=\langle x|\langle y| \rho | y\rangle |x\rangle$. We
have solve numerically the evolution using the exact
Hamiltonian, including both dimensions $x$ and $y$. We have
taken $\eta=0.5$, and $n=2$ in each dimension, and $\Omega=300\nu$
($\nu_x=\nu_y\equiv\nu$). The different figures correspond to
the state of the ion: (a) initial state; (b) after step (ii);
(c) after a time $\tau=\pi/(4\nu)$; (d) after a time
$\tau=\pi/(2\nu)$; (e) state after a time $\tau=3\pi/(4\nu)$;
(f) after a time $\tau=40 \pi/\nu$ (i.e. after 20 roundtrips).
As expected, the ion performs round trips without practically
modifying the structure of the wavepackets. Note that the
figures correspond to the diagonal part of the density operator,
and therefore one could have the ion in a statistical
mixture with exactly the same distribution. However, the fact
that the state is a pure state makes it possible to observe
quantum interferences, as it will be shown in the next
sections. Finally, in these figures it can be observed that
there is a small dephasing of the wavepackets [compare Figs.~4(c)
and (e)]. This is due to the fact that the pulses have a finite
duration, and during this time the wavepackets evolve slightly
[that is, conditions (\ref{condnut}) are not exactly fulfilled].
However, most part of this dephasing can be controlled, since
correspond to the free evolution with the harmonic oscillator
potential for a time equal to the pulse duration.

\section{Pure states versus statistical mixtures}

One of the crucial predictions of quantum mechanics is the
possibility of having coherent superpositions of certain states,
as it is the case for a Schr\"odinger cat state. Many of the
intriguing features of quantum mechanics rely on this
property. Thus, it would be highly desirable to have a way to
check experimentally whether the final state is a pure state of
the form (\ref{Schcat}) or an incoherent superposition
of the form
\begin{equation}
\label{incsup}
\rho \propto (|\eta\rangle_c \langle \eta | +
|-\eta\rangle_c \langle -\eta |) |e\rangle \langle e|.
\end{equation}
Similarly as for the pure state (\ref{Schcat}),
in this statistical mixture the ion is either to the right
(state $|\eta\rangle$) or to the left (state $-|\eta\rangle$) in
the $x$ axis.

Let us state this problem in a different (but equivalent) way.
Suppose we have our ion in the state (\ref{incsup}). Consider
the following experiment consisting of four steps:

\begin{description}

\item (i) We measure if the ion is to the right. This can be
done by using the quantum jump technique \cite{Na86} with a
laser focused on the right side only (after changing the state
of the ion in the right hand side from $|e\rangle$ to
$|g\rangle$, as indicated in Section III, i.e. using an
auxiliar laser propagating along the $z$ direction). If we do
not detect the ion, then we know that the ion is in the
left side, and its state is
\begin{equation}
|\Psi\rangle = |-\eta\rangle_c |e\rangle.
\end{equation}

\item (ii) Now we wait for a time $\tau=\pi/(2\nu)$, so that the state
of the ion will be $|i\eta\rangle|e\rangle$, and its position
will be centered around $x=0$.

\item (iii) We send two $\pi/2$ pulses along the $x$ direction,
the first propagating towards the negative values of $x$ while
the second propagating in the opposite direction. The state of
the ion after this pulse sequence can be easily calculated
using (\ref{evolpuls}), resulting in
\begin{equation}
|\Psi\rangle = K [
|e\rangle (|i\eta\rangle_c - |3i\eta\rangle_c)
-i |g\rangle  (|0\rangle_c + |2i\eta\rangle_c)].
\end{equation}

\item (iv) We measure the state of the ion using the quantum
jumps technique. Obviously, we will obtain that the probability
of measuring the ion in its ground state is $1/2$.

\end{description}

We could perform again the same experiment, but now in step (i)
we measure if the ion is to the left. If we do not detect the
ion, we will conclude that the state of the ion is
\begin{equation}
|\Psi\rangle = |\eta\rangle_c |e\rangle.
\end{equation}
Following the same steps as before (ii) and (iii), the state of the
ion will be
\begin{equation}
|\Psi\rangle = K
[e\rangle (|-i\eta\rangle_c - |i\eta\rangle_c)
-i |g\rangle  (|0\rangle_c + |-2i\eta\rangle_c)].
\end{equation}
Again, if we measure the state of the ion we will measure the
ground state with a probability $1/2$.

Summarizing, we could state that regardless the position of the
ion (i.e. both if it is to the left or to the right), after
performing steps (ii) and (iii), we will detect half of the
times the ion in the ground state, and the other half in the
excited state. However, if we take the state (\ref{Schcat}) and
calculate what happens after applying steps (ii) and (iii), we
will obtain the state
\begin{eqnarray}
|\Psi\rangle &=& K
|e\rangle (|-i\eta\rangle_c - |3i\eta\rangle_c) \nonumber\\
& &-i |g\rangle
(2|0\rangle_c + |-2i\eta\rangle_c+|2i\eta\rangle_c).
\end{eqnarray}
With this state, the probability of detecting the atom in the
ground state is 3/4!.

Obviously, there is a catch in the above argument. In order to
predict that whatever we have for the ion we will find it in
the ground state with probability 1/2, we have used a property
which is foreign to quantum mechanics, namely realism
\cite{Be87}. We all know that quantum mechanics is not a
realist theory, and therefore the apparent contradiction that
we have explained in the previous paragraph shows this fact.
State (\ref{Schcat}) is a state that can
only be described by quantum mechanics, and there is no
classical (realist) analog. Note that the state (\ref{incsup})
can be indeed described by a realist theory, and therefore if
we perform on it steps (ii) and (iii) we will detect the ion in
its ground internal state with probability 1/2. Therefore, the
experiment outlined in this Section [steps (ii) and (iii)]
allows one to distinguish between a pure state and a
statistical mixture, since the results for the probability of
detecting the ground state of the ion are different.

\section{Ion interferometer}

In this Section we analyze the possibilities of building a
Ramsey interferometer \cite{Ramseyint} based on the ideas we
have used to prepare superpositions of coherent states. We will
analyze the case in which laser pulses are used instead of
adiabatic passage, although both techniques can be applied with
similar results.

Consider a single ion in a trap after being laser cooled (via
sideband cooling) to its ground state $|g\rangle|0\rangle$. An
atom interferometer (in one--dimension) may be constructed as
follows:

\begin{description}

\item (i) Apply a $\pi/2$ pulse in the $x$ direction such that
the state of the ion becomes
\begin{equation}
|\Psi\rangle = \frac{1}{2} (|g\rangle |0\rangle_c - i |e\rangle
|i\eta\rangle_c).
\end{equation}
Then, wait for a time $\tau=\pi/(2\nu)$ so that the ion
wavefunction (in position representation) splits into two
different wavepackets
\begin{equation}
|\Psi\rangle = \frac{1}{2} (|g\rangle |0\rangle_c - i |e\rangle
|\eta\rangle_c).
\end{equation}
Note that as before, for small values of $\eta$ one can use
intermediate pulses to split the wavepacket further.

\item (ii) Apply a field to the wavepacket that is centered to
the right. For example, assume that we transform the state
$|e\rangle\rightarrow \cos(\alpha) |e\rangle - i\sin(\alpha)
|g\rangle$. In this particular case, the wavefunction will
become
\begin{equation}
|\Psi\rangle = \frac{1}{2} [|g\rangle |0\rangle_c - i
(\cos(\alpha) |e\rangle -i\sin(\alpha) |g\rangle)
|\eta\rangle_c].
\end{equation}
Then, wait again for a time $\tau=\pi/(2\nu)$ so that the ion
wavefunction (in position representation) is centered around
$\langle x \rangle = 0$
\begin{equation}
|\Psi\rangle = \frac{1}{2} [|g\rangle |0\rangle_c - i
(\cos(\alpha) |e\rangle -i\sin(\alpha) |g\rangle)
|-i\eta\rangle_c].
\end{equation}

\item (iii) Apply a $\pi/2$ pulse to the ion in the
opposite direction as in step (i). The state of the ion then will
become
\begin{eqnarray}
|\Psi\rangle &=& \frac{1}{2}
(|g\rangle \{[1-\cos(\alpha)]
|0\rangle_c - \sin(\alpha) |-i\eta\rangle_c \} \nonumber\\
& & -i |e\rangle \{[1+\cos(\alpha)]
|-i\eta \rangle_c - \sin(\alpha) |-2i\eta\rangle_c \}).
\end{eqnarray}

\item (iv) Measure the state of the ion using the quantum jumps
technique. One finds that the probability of measuring the
excited internal state is $P_e= \cos^2(\alpha/2)=1-P_g$. So,
depending on the phase $\alpha$ this probability varies as in
Ramsey spectroscopy.

\end{description}

In Fig.~5 we have plotted the probability of finding the ion in
the excited state as a function of the phase $\alpha$,
calculated numerically using the exact Hamiltonian. Here
$\eta=2.5$ and the  different curves correspond to
$\Omega/\nu=1,4,10$ and $100$ [solid, dashed, dotted and
dash--dotted lines, respectively]. In the strong excitation
regime (dash--dotted line), the visibility of the fringes is
nearly one. However, as the ratio $\Omega/\nu$ decreases, the
visibility is getting smaller and smaller, and even the curve
is distorted. The reason is that the condition (\ref{condnut})
is not satisfied anymore.

This kind of interferometer is very similar to the well--known
Ramsey spectroscopy. However the main difference is that the
phase shift here can be non--local, in the sense that one can
do something different to the different wavepackets, as it is
the case in atom interferometry \cite{Atomint}. Thus, a
technique like this could be useful to measure gradients of
fields, since the phaseshift on each wavepacket would depend on
its corresponding position. Apart from that, following the
lines of Subsection III.B one can generalize this to two
dimensions. In this case, one would have two wavepackets going
around a circle but in opposite directions. An interferometer
like this might be used to measure the Sagnac effect, where the
phase acquired by one of the wavepackets is different to that
of the other when the whole ion trap rotates at a given
frequency.

\section{Conclusions}

In this paper we have analyzed the interaction of a single
trapped ion with a laser beam in the strong excitation limit.
We have considered two situations: first, the ion interacts
with pulses of well defined area; secondly, the frequency of
the laser changes adiabatically. We have shown under which
limits these two situations give the same result regarding the
modification of the ion motion during the excitation.
Furthermore, Schr\"odinger cats \cite{note1} of the ion motion
can be prepared in an ion trap using both methods. We have
illustrated these methods with numerical calculations which
display the preparation of superpositions of coherent states
both in one and two dimensions. We have shown how one can check
the purity of these states, by relating this problem to the one
regarding the different predictions of (macro)realist's
theories and quantum mechanics. Finally, a trapped ion can be
used as an inteferometer by splitting the wavefunction and
looking at the interferences in the internal states.

The technique presented here can be easily generalized to
include the preparation of more general non--classical states,
including linear superposition of multiple coherent states with
the same amplitude and different phases. On the other hand, we
plan to study in detail the problem of how the Schr\"odinger
cat state dissipates in time in the presence of decoherence due
to spontaneous emission \cite{Ca85}.

In order to observe experimentally the behavior predicted in
this paper it is needed to fulfill the conditions given in
Section II, i.e. basically to work in the strong excitation
regime. This regime may be difficult to achieve, since one is
using an electric dipole forbidden transition and therefore the
laser intensity required to reach such a regime must be very
high. Alternatively, one could work with small trap frequencies
(i.e. out of the Lamb--Dicke limit), which would make it easier
to perform local observations. However, laser cooling to the
ground state of the trapping potential has not been observed
yet in this regime. One possible way to avoid this drawback is
to open the trap adiabatically (in order to decrease the trap
frequency, and the Lamb--Dicke parameter) once the ion has been
cooled in the Lamb--Dicke limit.

\section*{Acknowledgments}

J.F. Poyatos is supported by a grant of the Junta de
Comunidades de Castilla--La Mancha. Part of this work has been
supported by the Austrian Science Foundation.



\begin{figure}
\caption{
Level scheme of the internal transitions of a trapped
ion. Transition $|g\rangle \leftrightarrow |e\rangle$ is dipole
forbidden, whereas transition $|g\rangle \leftrightarrow
|r\rangle$ is allowed.
}
\end{figure}

\begin{figure}
\caption{
Real part of the density operator in momentum
representation $\langle p|\rho|p'\rangle$ in arbitrary units
[note that the axis are rescaled in terms of
$p_0=(m\nu/2)^{1/2}$], after the preparation of the
superposition state using laser pulses. Figures (a,b,c)
correspond to $\eta=0.5$ and $n=2$, whereas Figures (d,e,f)
correspond to $\eta=2.5$ and $n=0$. Here $\Omega/\nu$: $100$
(a,d), $10$ (b,e), and $1$ (c,f).
}
\end{figure}

\begin{figure}
\caption{
Real part of the density operator in momentum
representation $\langle p|\rho|p'\rangle$ in arbitrary units
[note that the axis are rescaled in terms of
$p_0=(m\nu/2)^{1/2}$], after the preparation of the
superposition state using adiabatic passage. Here,
$\Delta/\Omega=10$, $\Omega/\nu=100$ $\eta=0.5$, $n=2$, and:
(a) $\Omega\tau=40$; (b) $\Omega\tau=50$; (c) $\Omega\tau=60$;
(d) $\Omega \tau=70$.
}
\end{figure}

\begin{figure}
\caption{
Snap shots of the probability distribution in position
representation in two dimensions $\langle x|\langle y| \rho |
y\rangle |x\rangle$ in arbitrary units [note that the axis are
rescaled in terms of $x_0=y_0=1/(2 m nu)^{1/2}$]. Here
$\Omega/\nu=300$, $\eta=0.5$, $n=2$, and: (a) initial state;
(b) just after the superposition state preparation; (c) after
$\nu\tau=\pi/4$; (d) after $\nu\tau=\pi/2$; (e) after
$\nu\tau=3\pi/4$; (f) after $\nu\tau=40\pi$.
}
\end{figure}

\begin{figure}
\caption{
Excited state population $P_e$ as a function of the
phase $\alpha$ (see text for explanation). Here, $\eta=2.5$,
$\Omega/\nu$: 1 (solid line); 4 (dashed line); 10 (dotted line);
100 (dash--dotted line).
}
\end{figure}


\begin{references}
\bibitem{note1}
We use here the term macroscopic in the sense that the states
$|\alpha\rangle_c$ and $|-\alpha\rangle_c$ are distinguishible.
That is, if we shine resonant light on the ion we will observe
fluorescence at one of two possible locations in space, which
are separated by at least one wavelength (typically several
wavelengths). In other contexts, the term macroscopic is
reserved to denote states with a ``macroscopic'' number of quanta. In our
case, the states $|\alpha\rangle_c$ have typically 10 phonons. In
that sense, it would be more correct to use the term
``mesoscopic'' (see Ref.~\cite{Br92}).
\bibitem{note2}
Strictly speaking, condition (\ref{condnut}) is too
restrictive. It would suffice
$\nu_x\eta_x\sqrt{\overline{n}}\tau\ll 1$. In that case, one
can neglect the second term in Hamiltonian
(\ref{H_-}) during the interaction, and redefine the
detuning $\delta$, such that the Hamiltonian becomes $H=\nu_x
a^\dagger a + \frac{1}{2} \Omega \sigma_x$ (for $\delta=0$).
\end{references}
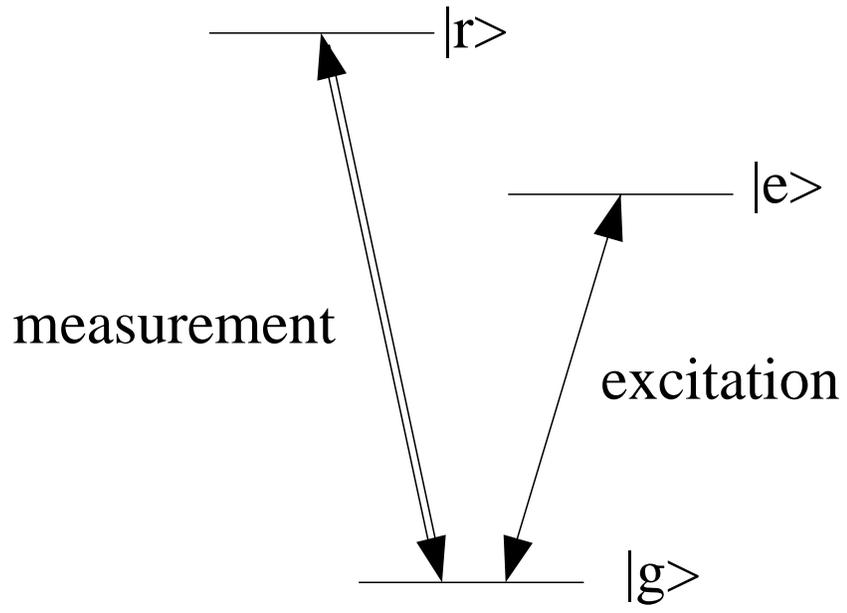
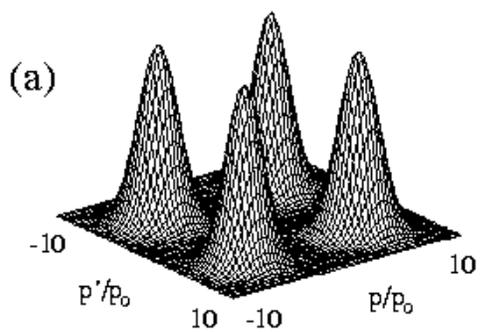
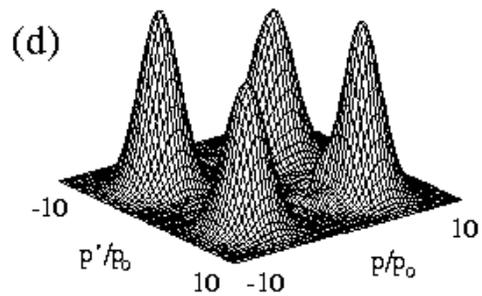
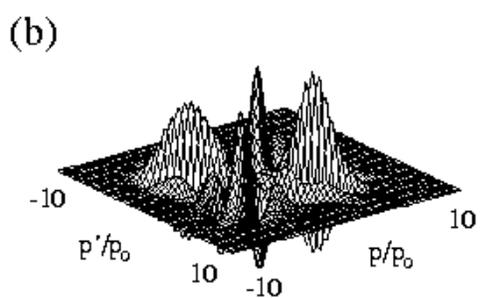
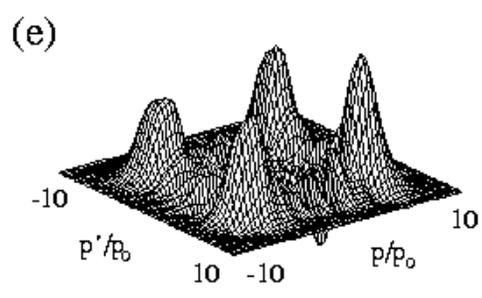
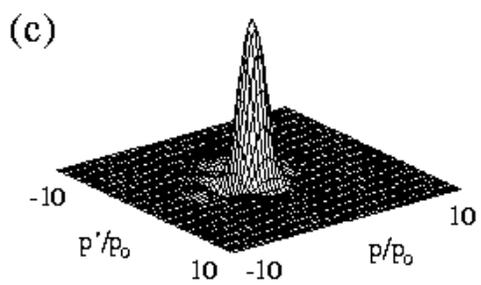
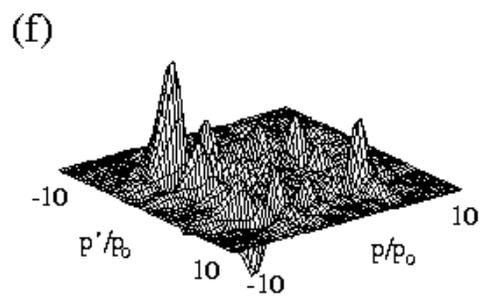
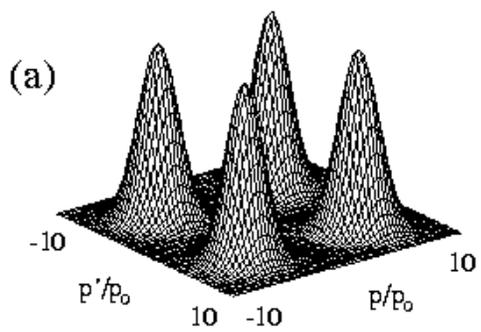
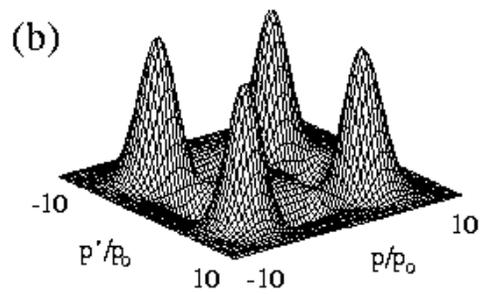
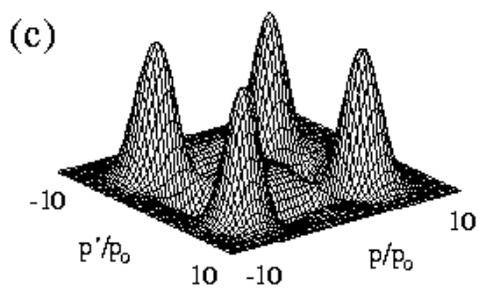
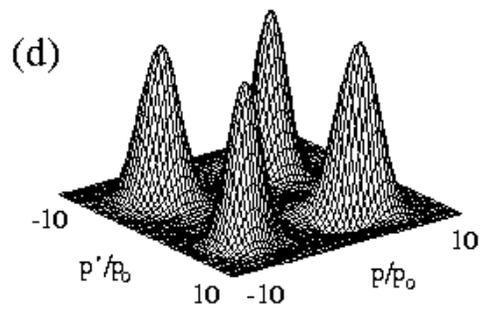
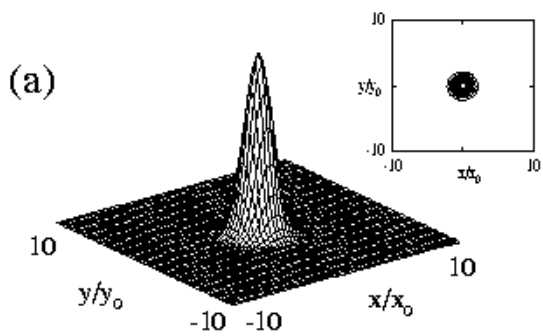
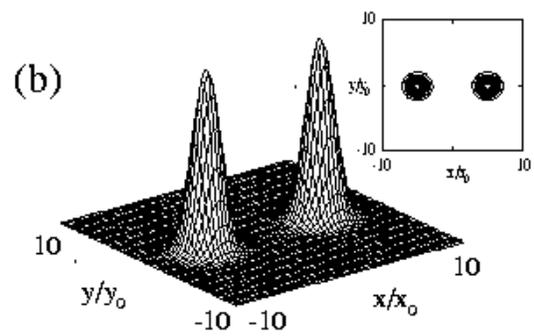
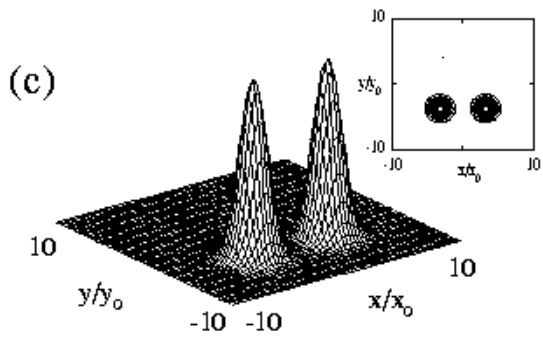
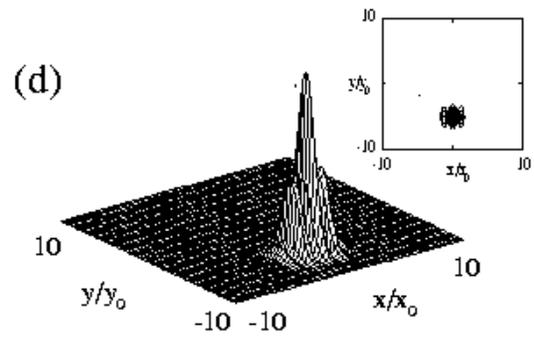
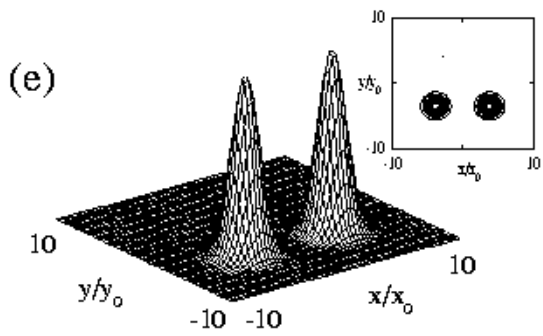
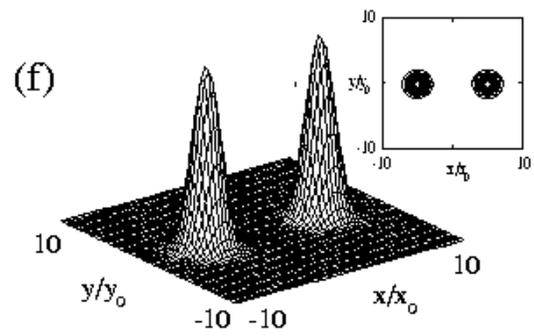
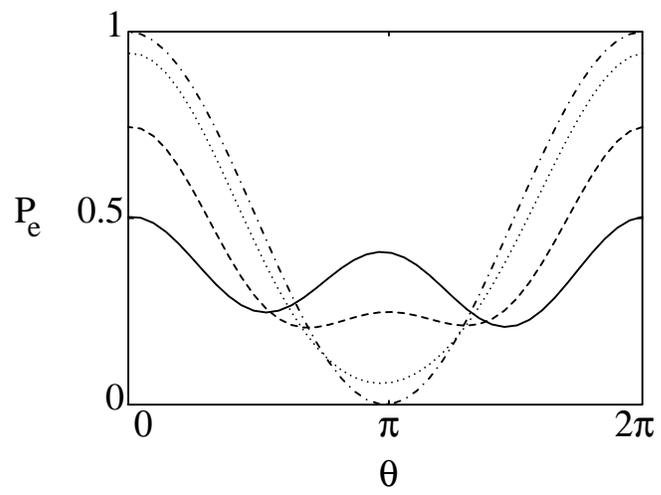
\end{document}